\documentclass[useAMS,usenatbib]{mnras}

\usepackage[english]{babel}
\usepackage{amsmath}
\usepackage{amssymb}
\usepackage{txfonts}
\usepackage{graphicx}

\newcommand{\ud}{\ensuremath{\mathrm{d}}}

\newcommand{\kB}{k_\mathrm{B}}
\newcommand{\Msun}{\ensuremath{\mathrm{M}_\odot}}

\title{Properties of Convective Oxygen and Silicon Burning Shells in Supernova Progenitors}

\author[C.~Collins et al.]{
  Christine~Collins$^{1}$\thanks{E-Mail: ccollins22@qub.ac.uk},
  Bernhard~M\"uller$^{2,1}$\thanks{E-mail: bernhard.mueller@monash.edu},
  Alexander~Heger$^{2,3,4}$\thanks{E-mail: alexander.heger@monash.edu}
\\
$^1${Astrophysics Research Centre, School
  of Mathematics and Physics, Queen's University
  Belfast, Belfast, BT7~1NN, United Kingdom} \\
$^2${Monash Centre for Astrophysics, School of
  Physics and Astronomy, Monash University, Victoria
  3800, Australia} \\
$^3${School of Physics \& Astronomy,
  University of Minnesota, Minneapolis, MN 55455, U.S.A.} \\
$^4${Department of   Astronomy, Shanghai Jiao-Tong University, Shanghai
  200240, P. R. China.} \\
}

\begin{document}

\label{firstpage}
\pagerange{\pageref{firstpage}--\pageref{lastpage}}

\maketitle

\begin{abstract}
Recent three-dimensional simulations have suggested that convective
seed perturbations from shell burning can play an important role in
triggering neutrino-driven supernova explosions. Since isolated
simulations cannot determine whether this perturbation-aided mechanism
is of general relevance across the progenitor mass range, we here
investigate the pertinent properties of convective oxygen and silicon
burning shells in a broad range of presupernova stellar evolution
models. We find that conditions for perturbation-aided explosions are
most favourable in the extended oxygen shells of progenitors between
about 16 and 26 solar masses, which exhibit large-scale convective
overturn with high convective Mach numbers. Although the highest
convective Mach numbers of up to 0.3 are reached in the oxygen shells
of low-mass progenitors, convection is typically dominated by
small-scale modes in these shells, which implies a more modest role of
initial perturbations in the explosion mechanism.  Convective silicon
burning rarely provides the high Mach numbers and large-scale
perturbations required for perturbation-aided explosions.  We also
find that about $40\%$ of progenitors between 16 and 26 solar masses
exhibit simultaneous oxygen and neon burning in the same convection
zone as a result of a shell merger shortly before collapse.
\end{abstract}


\section{Introduction}
\label{sec:intro}
It has long been recognised that the explosions of massive stars as core-collapse supernovae are inherently multi-dimensional phenomena \citep[for reviews see][]{janka_12,foglizzo_15}: In the modern version of the neutrino-driven explosion mechanism, multi-dimensional (multi-D) instabilities like convection \citep{herant_94,burrows_95,janka_96} and the standing accretion shock instability \citep{blondin_03,foglizzo_07} play a crucial role in
boosting the efficiency of neutrino heating sufficiently to revive the
shock; and for alternative scenario such as the magnetorotational
mechanism, the importance of multi-dimensional effects is even more
obvious.  The breaking of spherical symmetry is also crucial for
understanding many features of the observable transients and the
compact and gaseous remnants: After shock revival, the asymmetries in
the explosion determine the kick and spin of the neutron star
\citep{janka_94,burrows_96,fryer_04a,scheck_06,wongwathanarat_10b,rantsiou_11}. Further
multi-dimensional instabilities come into play as the shock propagates
through the outer shells of the progenitor out to the stellar envelope
and give rise to mixing as already recognised in 
the 1970s \citep{falk_73,chevalier_76}.

Recently, multi-dimensional effects during the final stages of
\emph{presupernova} evolution have garnered particular interest in
supernova theory. It has been suggested that the asymmetries seeded by
convective shell burning could play an important role in tipping the
scales in favour of shock revival in neutrino-driven explosions
\citep{couch_13,mueller_15a} by boosting the violent non-spherical
motions in the gain region behind the shock upon the infall of those
shells. \citet{mueller_15a} showed that the physical mechanism behind
``perturbation-aided'' explosions involves the conversion of the
initial convective velocity perturbation into large density and ram
pressure perturbations at the shock during the infall, which distort
the stalled accretion shock and foster the development of large
buoyant bubbles and fast non-radial flows in the gain region.  They
also established qualitatively that the crucial parameters for this
``perturbation-aided mechanism'' are the initial convective Mach
number and the angular scale of the convective eddies in the
progenitor; faster convective flow and larger eddies are more
conducive to perturbation-aided explosions. Subsequent studies have
attempted to put these qualitative trends on a more quantitative
footing \citep{mueller_16c,abdikamalov_16}.

The most direct way to gauge the role of convective perturbations in
the progenitor in the neutrino-driven mechanism is to initialise
supernova simulations properly by simulating at least the last few
turnover timescales of shell convection in multiple dimensions. After
a long series of two- and three-dimensional simulations of oxygen and
silicon shell burning during earlier phases
\citep{arnett_94,bazan_94,bazan_98,asida_00,kuhlen_03,meakin_06,meakin_07,meakin_07_b,arnett_11,jones_17},
attempts to evolve convective shells up to the onset of collapse in 3D
were first made by \citet{couch_15} for silicon burning in a $15
\,\Msun$ star, and for oxygen burning by \citet{mueller_16c} and
\citet{mueller_16b} for $18 \,\Msun$ and $12.5 \,\Msun$
progenitors. Follow-up simulations of the collapse and post-bounce
phase of the ensuing supernova with different methodology have yielded
very different results ranging from a  small impact on the
neutrino heating conditions \citep{couch_15} to a large qualitative
difference in \citet{mueller_17}, where the perturbations prove crucial
for triggering shock revival.

Such a diversity of effect sizes is not unexpected since the shell
configuration and shell burning rates vary significantly among
supernova progenitors. This implies that a few isolated multi-D
simulations are not sufficient for determining whether convective seed
perturbations generically play a major role in the supernova explosion
mechanism. In addition to more extensive studies of the perturbation-aided
mechanism in 3D, a more systematic investigation of the relevant
parameters of shell convection across a large number of progenitors
is needed, both to extrapolate the findings of the limited set of 3D studies and to help better target the 3D simulations towards interesting stellar models.

In this paper, we undertake such a study for the first time.  Based on
the realisation that one-dimensional stellar evolution models can be
used to estimate both the violence of convection and the geometry of
convective eddies even during the very dynamical phase right before
the onset of collapse \citep{mueller_16c}, we investigate the
systematics of the convective Mach number and the geometry of the
innermost active burning shells for a set of 2,353 one-dimensional
stellar evolution models \citep{mueller_16a}. 

With this survey of
progenitor models, we can shed light on the following questions
concerning the perturbation-aided mechanism: Are pre-collapse
perturbations generically strong enough to efficiently aid
neutrino-driven shock revival? What mass ranges are particularly
promising for perturbation-aided explosion?  Are perturbations in the
oxygen or silicon shell more promising for achieving shock revival? Are there unexplored scenarios that need to be simulated in 3D because some assumptions of the mixing-length approach in 1D stellar evolution models break down?
Naturally, we cannot answer these questions in a definitive manner by
merely considering spherically symmetric stellar evolution models;
detailed 3D models of the last episodes of shell burning and the
subsequent supernova explosion remain indispensable. One of the
primary purposes of the present study is to provide guidance for such
simulations by allowing a more targeted selection of stellar progenitor
models according to the properties of the innermost convective shells.

Our paper is structured as follows: In Section~\ref{sec:models}, we
briefly describe the stellar evolution models used in this study and
review how convective velocities and eddy scales can be estimated from
1D models. In Section~\ref{sec:results}, we investigate systematic
trends in convective Mach number and eddy scale in different shells
and discuss how these are related to variations of stellar structure.
We then present tentative estimates for the impact of convective
perturbations on ``explodability'' across the stellar mass range. We
conclude in Section~\ref{sec:conclusions} by discussing the
implications of our results for future studies of the
perturbation-aided explosion mechanism.

\section{Input Models and Theoretical Background}
\label{sec:models}

\subsection{Stellar Evolution Models}

We consider a set of 2,353 solar-metallicity non-rotating progenitor
models with zero-age main sequence (ZAMS) masses between $9.45
\,\Msun$ and $35\,\Msun$ that have been computed with the stellar
evolution code \textsc{Kepler} \citep{weaver_78,heger_10}.  This is an
extension of the set of progenitors presented in \citet{mueller_16a}.
In these models, at $35\,\Msun$, the Si core mass becomes comparable
to the maximum current lower limit for the maximum baryonic neutron
star mass \citep{demorest_10,antoniadis_16}, and peturbation-aided
explosions by the neutrino-driven mechanism therefore become unlikely.
For this reason we do not present more massive progenitors.  For high
masses, however, mass loss may become important and reduce the core
mass, which may affect the final outcome.  We assume that the core
structure of the star is not much affected by mass loss as long as
some hydrogen envelope is left.  Mass loss in massive stars remains somewhat
uncertain, and is not the topic of this study.

The models were calculated using the standard treatment of mixing in
\textsc{Kepler}, i.e., convective mixing according to mixing-length
theory in Ledoux-unstable regions, semiconvection according to
\citet{weaver_78}, and thermohaline mixing following \citet{heger_05}.
These models differ from the compilation by
\citet{sukhbold_14,sukhbold_16} by updated neutrino physics that shift
some of the features of late at evolution in the $15\,\Msun \ldots
20\,\Msun$ region by $1\,\Msun \ldots 2\,\Msun$.  For numerical reasons,
convective regions are bounded by one zone of overshooting with an
efficiency similar to semiconvection.  For the initial composition, we
use the initial (in contrast to present-day) solar abundances of
\citet{asplund_09}.

Following the extant studies of perturbation-aided explosions from 3D
progenitors models \citep{couch_15,mueller_16b,mueller_17}, we focus
only on convection driven by Si or O burning. This is motivated
by several considerations.

The O and Si shell
are most likely to provide significant perturbations that can aid
neutrino-driven explosions, and there is theoretical and observational
evidence to suggest that shock revival occurs either during the infall
of the Si or O shell. Both in multi-D supernova simulations
\citep{suwa_14,summa_16} and in parameterised numerical
\citep{ugliano_12,ertl_15} and analytic \citep{mueller_16a},
1D models, shock revival is often associated with the
infall of the Si/O shell interface, though there is also a number of
cases where the explosion is initiated later during the accretion of
the O shell.  Delaying the explosion until the Ne or C shell reach the
shock could lead to tensions with the observed distribution of neutron
star masses \citep{schwab_10,oezel_12,oezel_16}.
Moreover, the
assumption of a mass cut close to the Si/O interface has been found to
lead to adequate matches of the population-integrated core-collapse
supernova nucleosynthesis and the solar abundance pattern
\citep{woosley_02}. We do, however, consider merged O/Ne/C shells in
which both O and Ne burning are active (Section~\ref{sec:merger}).

Whether shock revival occurs during the infall of the O shell or the
Si shell is less clear, and our study attempts to clarify this
further.  Simulations as well as supernova nucleosynthesis provide
circumstantial evidence that shock revival generally does not occur in the Si
shell.  Due to the significant neutron excess, explosive burning in
the Si shell would lead, for example, to higher Ni/Fe ratios than
required for chemical evolution \citep{arnett_96} and than
observed for most core-collapse supernovae. There are, however,
exceptions among the observed core-collapse events that might be
explained by ejection of material from the Si shell
\citep{jerkstrand_15b}. The fact that the Si/O shell interface often
acts as the trigger for shock revival in numerical models has already been mentioned, but
on the theoretical side there are also examples \citep{mueller_12b} of
explosions that already develop before the infall of the shell
interface (even though such cases may be problematic because of
modelling assumptions such as the restriction to axisymmetry). Furthermore, the point of
shock revival and the ``mass cut'' are not strictly associated with each other in multi-D models \citep{harris_17,wanajo_17}, and early shock revival
in the Si shell is therefore not necessarily
in conflict with nucleosynthetic constraints.

Whether the pre-collapse perturbations from convective Si or O burning
are dynamically more important for shock revival is also highly
relevant for 3D models of the last minutes of shell burning. Due to
the quasi-equilibrium nature of Si burning \citep{bodansky_68,woosley_73,hix_96}, rigorous 3D
simulations of convective Si shells are technically more
challenging. Although the basic behaviour of Si shell burning may be
captured qualitatively by small networks (as in \citealt{couch_15}),
correctly treating the gradual shift of abundances from Si and S to
the iron group and the deleptonisation requires large networks. If Si
shell burning needs to be taken into account when constructing 3D
supernova progenitor models, this implies a significant increase in
computational cost and complexity.

For these reasons, we consider both convective Si and O shell
in our progenitors in this study. There are some cases where
there is more than one active convective shell driven by these burning processes
(e.g., when a partially unburnt layer of O below the main O layer reignites
shortly before collapse due to the contraction of the core.
In these cases, we consider only the shell with the maximum
convective velocity as the one that is most likely to trigger
a perturbation-aided explosion.

\subsection{Estimating Convective Velocities and Eddy Scales}
We estimate convective velocities and Mach numbers using mixing-length
theory
(MLT; \citealp{biermann_32,boehm_58}) and convective eddy scales based on the
geometry of the convective shells assuming that the largest eddies
stretch across the entire convective zone and are of similar vertical
and horizontal extent. Simulations of advanced, neutrino-cooled
burning stages \citep{kuhlen_03,arnett_09,mueller_16c,jones_17}
suggest that the bulk properties of the 3D velocity field are reasonably
well captured by these simple estimates if the dimensionless
coefficients in the MLT equations are chosen appropriately.

\subsubsection{Convective Velocity and Mach Number}
The convective velocity $v_\mathrm{conv}$ in MLT can be expressed by
\begin{equation}
  \nonumber
\label{eq:vconv}
v_\mathrm{conv}
=
\alpha_1 \Lambda_\mathrm{mix} \omega_\mathrm{BV}
\end{equation}
in terms of a non-dimensional coefficient $\alpha_1$,
the mixing length $\Lambda_\mathrm{mix}$, and the
Brunt-V\"ais\"al\"a frequency $\omega_\mathrm{BV}$,
which, in turn, is given by
\begin{equation}
\omega_\mathrm{BV}=\sqrt{g \left(\frac{\ud \ln \rho}{\ud r}- \frac{1}{\Gamma}\frac{\ud \ln P}{\ud r }\right)}\;.
\end{equation}
Here $g$ is the local gravitational acceleration, $r$ is the radial coordinate, and $\rho$,
$P$, and $\Gamma$ are the density, pressure, and adiabatic index, respectively.

Following \citet{mueller_16c}, we set the mixing length
to one pressure scale height,
\begin{equation}
\Lambda_\mathrm{mix}=\frac{P}{\rho g},
\end{equation}
and set $\alpha_1=1$, which resulted in good agreement with the
convective velocities in their 3D simulation of oxygen shell burning.

A potential problem with the MLT estimate arises right before the
onset of collapse when the convective turnover timescale and the
contraction timescale become similar and convection can no longer
adjust to the acceleration of nuclear burning during collapse.
\citet{mueller_16c} found, however,
that this freeze-out of convection is still captured well by
1D stellar evolution models employing time-dependent MLT so that
the convective velocities from our \textsc{Kepler} models
remain good first-order estimates up to the onset of collapse. For a discussion of possible uncertainties
in our predictions due to the use of time-dependent MLT,
we refer the reader to Section~\ref{sec:mlt_uncertainties}

\citet{mueller_15a} showed that
the relevant quantity for perturbation-aided explosion
is the convective Mach number, $\mathrm{Ma}=v_\mathrm{conv}/c_\mathrm{s}$
where $c_\mathrm{s}$ is the sound speed, in the progenitor
rather than the convective velocity. 
This is because the Mach number roughly reflects
the ratio of the advective crossing timescale
$t_\mathrm{cross}=r/v_\mathrm{conv}$ 
and the free-fall timescale $t_\mathrm{ff}$, which determines
the density and ram-pressure perturbations at the shock after collapse.
For the sake of simplicity,
we therefore compute $c_\mathrm{s}$ assuming a constant adiabatic
index of $\Gamma=4/3$,
\begin{equation}
\label{eq:maconv}
\mathrm{Ma}=\frac{v_\mathrm{conv}}{c_\mathrm{s}}
\approx \sqrt{\frac{3 \rho v_\mathrm{conv}^2}{4 P}},
\end{equation}
which is a very good approximation for the convective burning shells
in question. Moreover, the relation between the pre-collapse Mach
number and the perturbation is not so tight as to warrant
the use of the exact value of $\Gamma$; other approximations
limit our analysis and conclusions much more seriously.

\subsubsection{Eddy Scale and Determination of Shell Boundaries}
3D simulations of advanced shell burning stages show that the size of
the largest eddies and the peak of the turbulent
energy spectrum are determined by the depth of the convection zone
\citep{arnett_09,mueller_16c}, which roughly corresponds
to the wavelength of the most unstable mode in the linear regime
 \citep{chandrasekhar_61,foglizzo_06}. The dominant
angular wave number, $\ell$, is therefore given by
\begin{equation}
\label{eq:lconv}
\ell= \frac{\pi (r_+ + r_-)}{2 (r_+ - r_-)},
\end{equation}
where $r_-$ and $r_+$ are the radius of the inner and outer boundary
of a convective zone. For a given convective shell, we determine $r_-$
and $r_+$ by locating the maximum of $v_\mathrm{conv}$ within
that shell and then identifying the interval around that
point in which $v_\mathrm{conv}$ exceeds $5\%$ of that maximum value.
This procedure was adopted to separate convective zones that lie
directly adjacent to each other, which is very often the 
case for the O and C shells. 

\begin{figure}
\includegraphics[width=\linewidth]{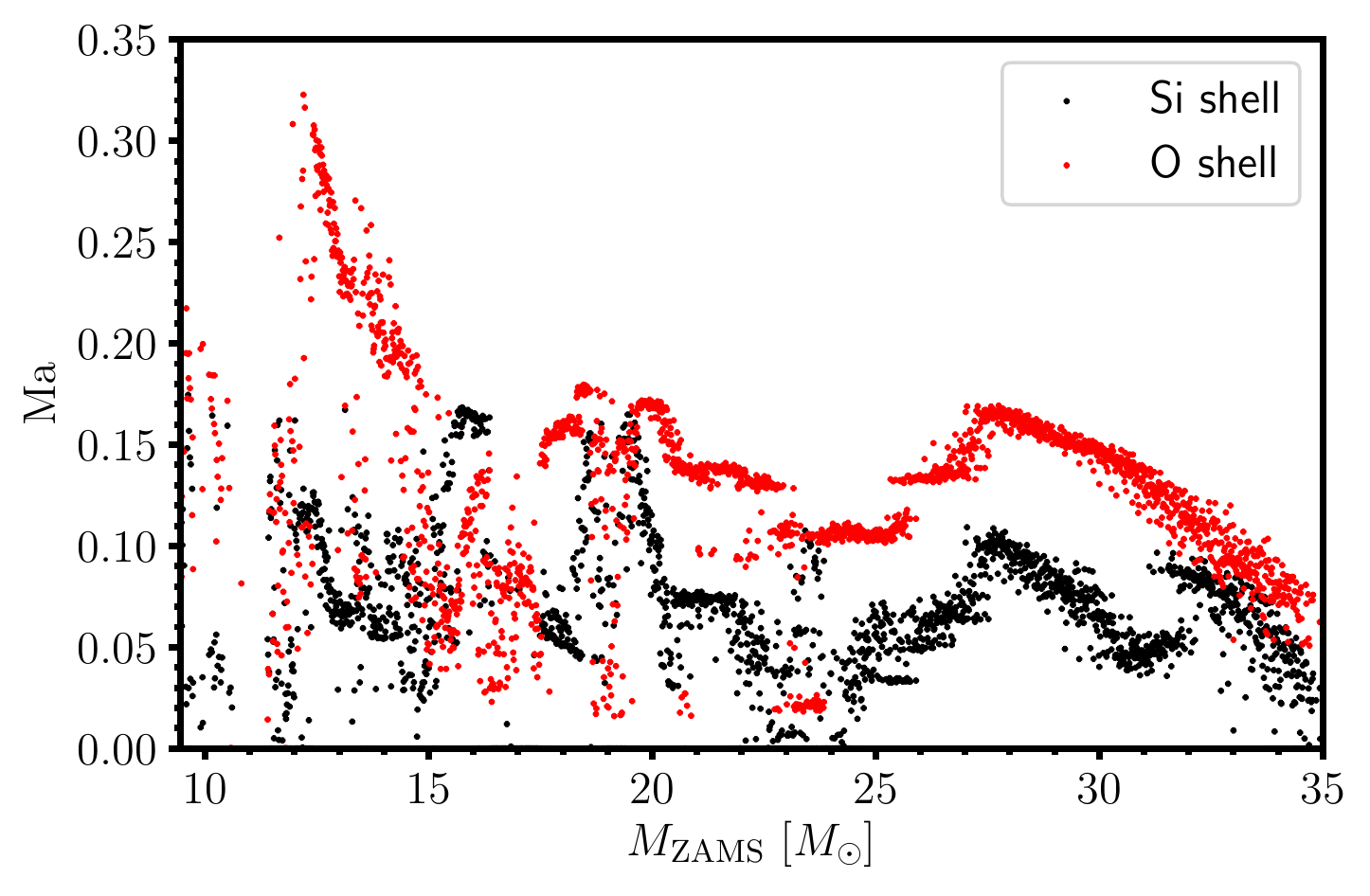}
\caption{Maximum convective Mach number
$\mathrm{Ma}$ in the Si (black)
and O (red) shell in all progenitor models at the onset of collapse.
Note that there is  a trend towards higher
$\mathrm{Ma}$ in the O shell
in the less massive progenitors,
and that the convective Mach number
in the O shell is generally higher
than in the Si shell.
\label{fig:mach}}
\end{figure}

\begin{figure}
\includegraphics[width=\linewidth]{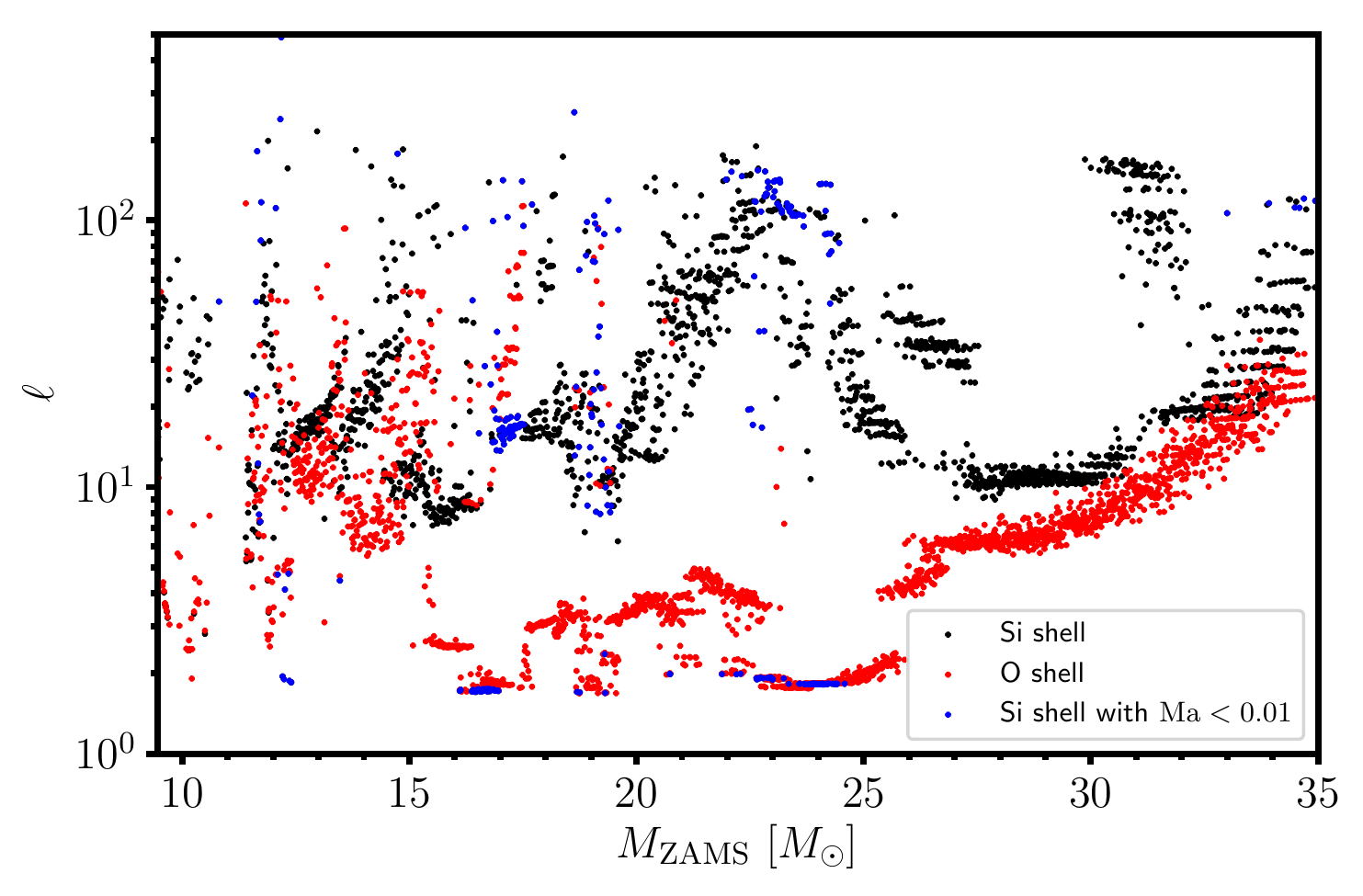}
\caption{Predicted angular wave number $\ell$ of the dominant convective mode
in the Si shell (black) and
O shell (red) for all pre-supernova
models. Si shells with 
very weak convection
($\mathrm{Ma}<0.01$ are shown in blue). Note that large-scale
modes with $\ell\lesssim 5$ are largely
confined to the O shells of progenitors in the
mass range between $16 \,\Msun$
and $26 \,\Msun$.
\label{fig:ang}}
\end{figure}

\section{Results}
\label{sec:results}

\subsection{Systematics of Convective Mach Numbers
at Collapse} The convective Mach numbers in the Si shell
and the O shell are shown in Figure~\ref{fig:mach} for all our
progenitor models. We note that we find higher values
(up to $0.3$) than
sometimes quoted in the literature for Si and O shell burning based
on 1D stellar evolution models. This is partly
due to our choice of dimensionless coefficients in
Equation~(\ref{eq:vconv}), which have been calibrated
to recent 3D simulations. Moreover, quoted values
of just a few $0.01$ for $\mathrm{Ma}$ often refer
to earlier stages of shell burning, and there is
a considerable increase in convective velocities
in the last few minutes prior to collapse as the
shells follow the contraction of the Fe core and become
hotter (see, e.g., Figure~6 in \citealt{mueller_16c}).

Clear systematic trends in convective Mach number are evident.
Convective Mach numbers in the O shell peak in the
mass range of $\mathord{\sim} 11.5\,\Msun\ldots 13\,\Msun$, where they
reach values of up to $0.3$, and generally decline with progenitor
mass. In the vast majority of progenitors O burning drives strong
convection with Mach numbers higher than $0.1$. There is, however, a
non-negligible fraction of progenitors that lack a strong active O shell
at collapse and show only modest or weak convective activity.  This is
particularly true of the window between $14.5 \,\Msun$ and $17.5
  \,\Msun$ in ZAMS mass. Above $20 \,\Msun$, outliers with weak convection in
the O shell become rare.

Convective Mach numbers in the Si shell are typically significantly smaller and lie mostly in the range $\mathrm{Ma}=0.05\ldots 0.1$.  $\mathrm{Ma}$ is smaller than $0.1$ in $79\%$ of our models (assuming IMF weighting with a Salpeter IMF $\propto M^{-2.35}$ as in the rest of this paper).  In $14\%$ of the progenitors,  Si shell convection is practically absent with $\mathrm{Ma}<0.01$.

Thus, O burning generally drives stronger convection with Mach numbers higher by a factor of $2\ldots 3$ for the majority of models; the convective Mach number in the O shell exceeds that in the Si shell in $88\%$ of our models.  Exceptions include small windows around  $15.5 \,\Msun$ and   $18\,\Msun\ldots 20\,\Msun$ in ZAMS mass, where the convective Mach numbers in the Si shell reach up to $0.15$ and equal or exceed those in the O shell.  

\subsection{Shell Geometry}
\label{sec:ShellGeo}
The predicted dominant angular wave numbers of convection in the Si and
O shell are plotted in Figure~\ref{fig:ang}.

Small angular wave numbers, $\ell \lesssim 5$, i.e., thick convective
shells, that are favourable for perturbation-aided explosions are
found predominantly for O shells in progenitors between $15 \,\Msun$
and $26\,\Msun$ and a small number of progenitors around $10\,
  \Msun$, and are rare outside this mass range.  It is noteworthy
that the O shells with high convective Mach numbers in low-mass
progenitors are, for the most part, rather thin and typically have
$\ell \sim 10$, although there is considerable scatter below $15
\,\Msun$.

Active Si shells prone to low-$\ell$ modes in convection are rare.  Only $7\%$ of the pre-supernova models have $\ell\leq5$ in the Si shell, and $84\%$ have $\ell > 10$.  Moreover, the Si shells with $\ell\leq5$ invariably have weak convection with $\mathrm{Ma}<0.01$.  Si shells with strong convection, i.e., $\mathrm{Ma}\gtrsim0.1$, typically have $\ell \sim 10$.  In $80\%$ of the models, we expect that Si shell burning is characterised by eddies of smaller angular scale than O shell burning at the onset of collapse.

\subsection{Structural Reasons for Variations in Shell Properties}

\subsubsection{Effect of Core Size and Degree of Shell Depletion for O Shells} 

\begin{figure}
\includegraphics[width=\linewidth]{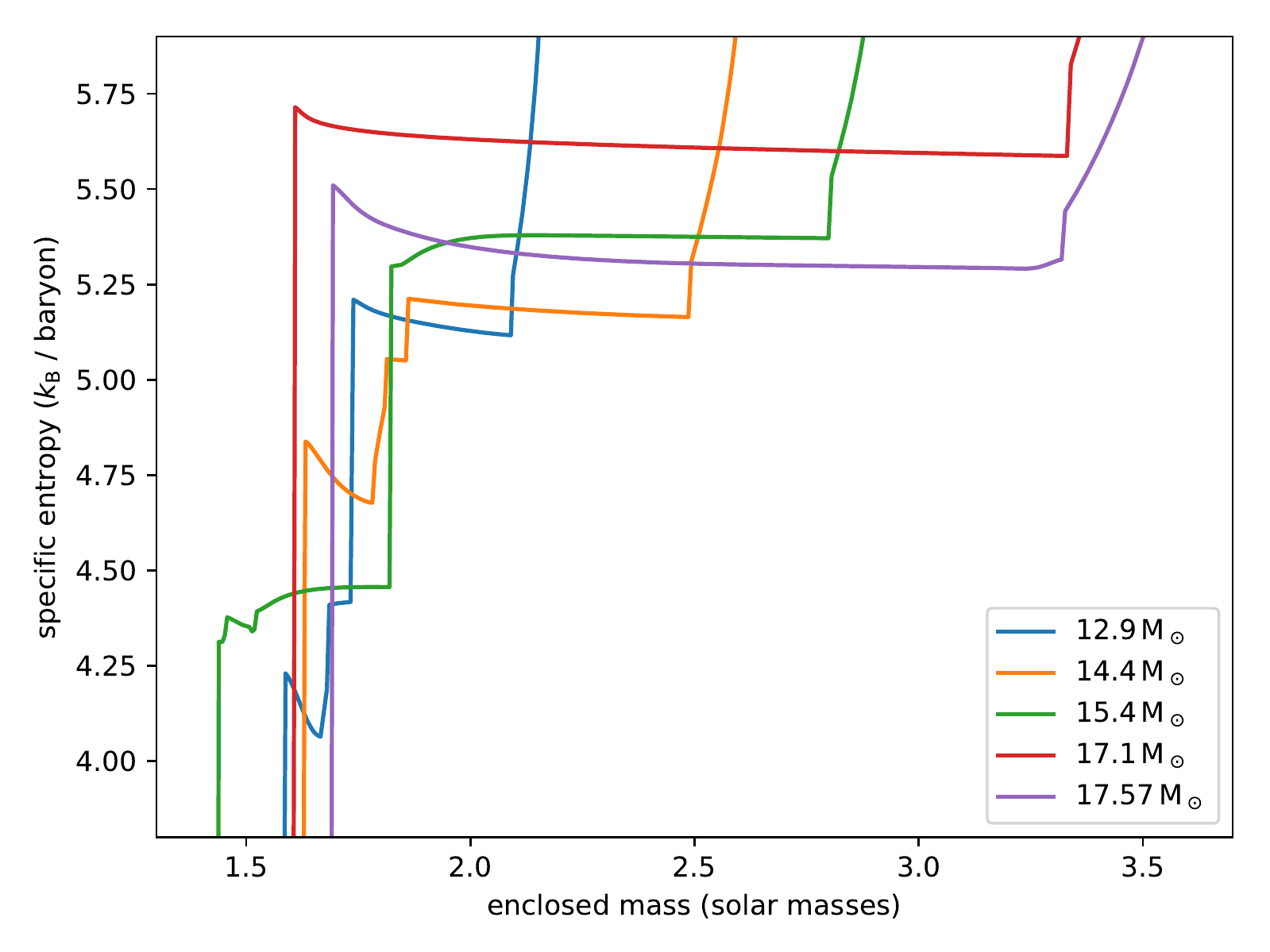}
\caption{Specific entropy at core collapse for different initial masses.  One can observe a large rise in entropy toward the bottom of the convective oxygen shell between $m=1.5\,\Msun$ and $1.7\,\Msun$, except for the $15.4\,\Msun$ model, where there is only a tenuous convective shell with weak burning.
\label{fig:S}}
\end{figure}

Negative entropy gradients, i.e., entropy inversions, is what drives the motion in convective regions.  Usually, in the stellar interior convection is quite efficient and one would expect almost flat entropy gradients inside convection zones.  Figure~\ref{fig:S} shows the entropy profile in the O shell for a select set of models.  The steep inversions toward the bottom of the O shell in the model is what drives the fast motions reported here for the collapsing O shells.  This is particularly apparent for the extreme case of the $17.57\,\Msun$ star.

The trends mentioned in Section~\ref{sec:ShellGeo} cannot be reduced to a uniform underlying principle of stellar structure.  Even the variations in the convective Mach number in the O shell, which exhibit a relatively clear decreasing trend from $12\,\Msun$ to $25\,\Msun$ in ZAMS mass, are the result of a complex interplay between the varying mass and radius of the Si core, the shell entropy and density, and different degrees of shell depletion.

\begin{figure}
\includegraphics[width=\linewidth]{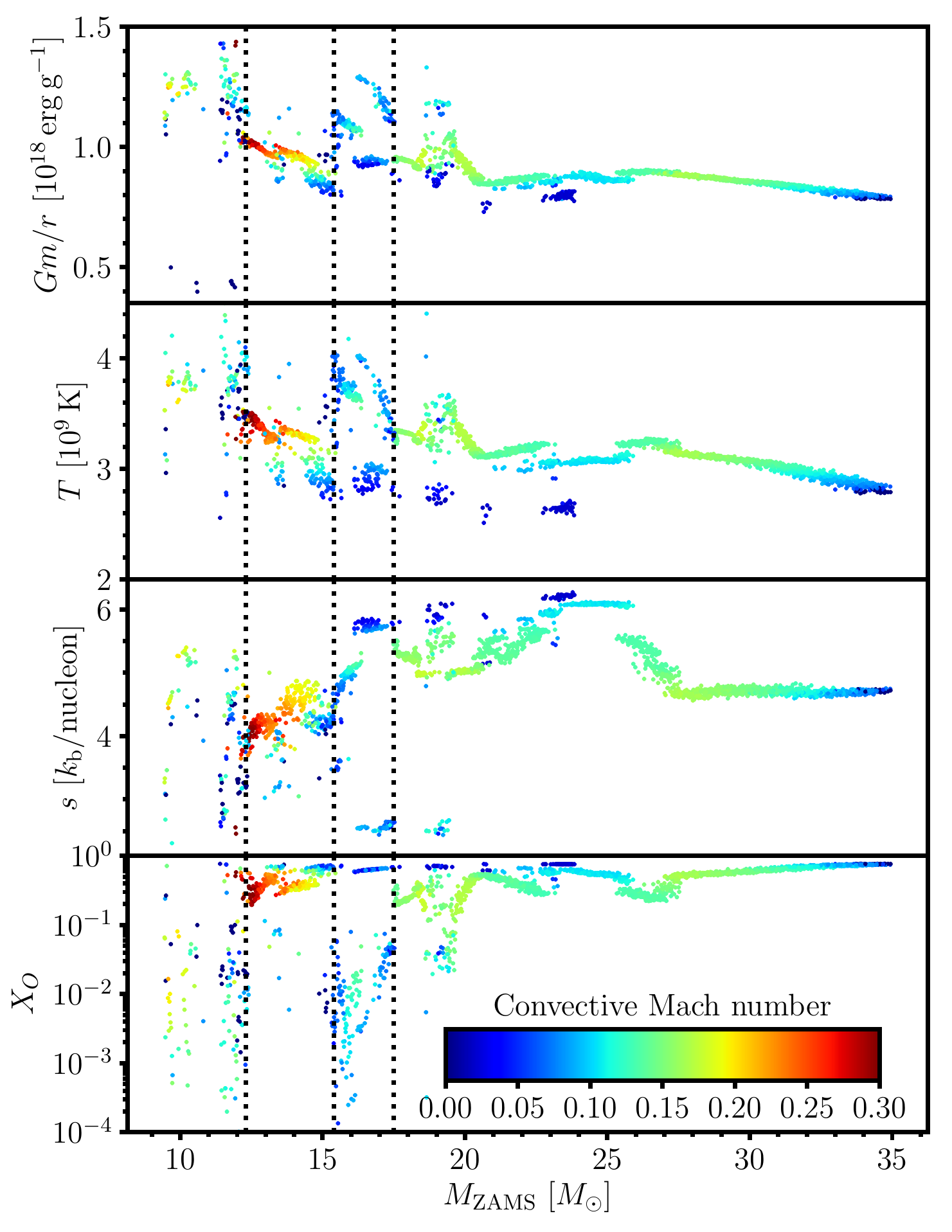}
\caption{Key properties of convective O shells as a function of ZAMS mass $M_\mathrm{ZAMS}$.  From top to bottom: gravitational potential $Gm/r$ of the Si core, and temperature, $T$, specific entropy, $s$, and O mass fraction $X_\mathrm{O}$ at the base of the O shell.  The maximum convective Mach number in the shell is color-coded. Vertical dashed lines denote boundaries between important regimes, such as the emergence of sufficiently high $X_\mathrm{O}$ for vigorous convection above $12.3\,\Msun$, and the region of slow O shell convection between $15.4\,\Msun$, and $17.4\,\Msun$. 
\label{fig:shells}}
\end{figure}

This is illustrated in Figure~\ref{fig:shells}, which shows the (absolute value of the) gravitational potential, $Gm/r$,\footnote{At the location of the oxygen shell this is a good proxy, typically within $20\% \ldots 30\%$, for the actual gravitational potential, $\left|\phi(r)\right|=Gm/r+\int_r^R Gm(r^\prime)/(r^\prime)^2\,\mathrm{d}r^\prime$, where $R$ and $M$ are the radius and total mass of the stars.
Moreover, $Gm/r$ is actually correlated more tightly
with shell temperature than $\phi$.} of the enclosed mass $m$ at the base of the O shell at radius $r$, and the temperature $T$, entropy $s$, and oxygen mass fraction $X_\mathrm{O}$ at the base of the shell.

Except for the intervals below $\mathord{\sim}12.5\,\Msun$ and
$15.4\,\Msun\ldots 17.5\,\Msun$, the convective Mach number correlates
visibly with the depth of the gravitational potential at the base of
the O shell and with the shell temperature, which is close to $\kB
T=0.3\,Gm m_\mathrm{N}/r$ (where $\kB$ and $m_\mathrm{N}$ are the
Boltzmann constant and the nucleon mass) and tends to be higher for
more compact Fe-Si cores with smaller mass, higher degeneracy, and
smaller radius.  The key here is that the \emph{effective}
  dependence of the core radius on the core mass is an inverse one and
  is relatively steep; thus less massive cores generate a
  \emph{deeper} potential well than more massive ones. Since the O
burning rate depends strongly on temperature ($\propto T^{25\ldots
  30}$), this trend towards hotter shells around smaller cores
explains the bulk of the variation in $\mathrm{Ma}$ in the intervals
between $12.3\,\Msun$ and $15.4\,\Msun$ and $>17.5 \,\Msun$ in ZAMS
mass.

The dependence of the O burning rate and the convective Mach number on temperature is, however, modified by variations in shell entropy and O mass fraction, in particular for ZAMS masses in the region $\mathord{<}12.3\,\Msun$ and $15.4\,\Msun \ldots 17.5\,\Msun$.  In these mass ranges we encounter clusters of progenitors with high temperatures in the O shell that only exhibit weak convective activity.  This is due to the fact that oxygen is close to depletion in these shells already (bottom panel of Figure~\ref{fig:shells}), so that the nuclear burning rate ($\propto X_\mathrm{O}^2$) is low despite high shell temperatures.  This partly explains the prevalence of weak convection in the O shell in the interval $15.4\,\Msun\ldots 17.5\,\Msun$, but there is also another branch of evolutionary channels in this range that has slow O shell convection for a different reason, namely low maximum shell temperatures of $\mathord{\sim} 3 \times 10^9 \,\mathrm{K} $ that occur in some merged O/Ne/C shells (see Section~\ref{sec:merger}).  

It is noteworthy that higher convective Mach numbers in the O shell tend to be correlated with lower shell entropies (except for the case of strongly depleted shells discussed above).  The low shell entropies in models with strong convection reflect the fact that O burning occurs deeper in the gravitational potential and hence at higher temperatures also earlier on in the life of the O shell.  The shell entropy can be understood as a tracer of the conditions that hold while nuclear energy generation and neutrino cooling still balance each other.  In a one-zone model, the condition of balanced power \citep{woosley_72}
\begin{equation}
\dot{\epsilon}_\mathrm{nuc} \sim \dot{\epsilon}_\nu,
\end{equation}
for the nuclear energy generation
rate 
$\dot{\epsilon}_\mathrm{nuc}$ and the neutrino 
cooling rate $\dot{\epsilon}_\nu$
leads to 
\begin{eqnarray}
\rho T^{\alpha} X_\mathrm{O}^2 &\sim& C \rho^{-1} T^\beta \\
\rho &\propto& X_\mathrm{O}^{-1} (T^{(\alpha-\beta)/2})
\end{eqnarray}
if we assume a power-law dependence for
$\dot{\epsilon}_\mathrm{nuc}$ and  $\dot{\epsilon}_\nu$
with $\alpha \sim 27$ for O burning and $\beta \sim 9$
in the regime where neutrino cooling is dominated by
the pair process. If the dominant contribution
to the entropy comes from photon radiation,
the entropy maintained by the shell is
\begin{equation}
s \propto \frac{T^3}{\rho}
\propto X_\mathrm{O} T^{3-(\alpha-\beta)/2}
\propto X_\mathrm{O} T^{-6},
\end{equation}
as long as the condition of balanced power holds.
Since one finds $k T/m_\mathrm{N}\mathrm{\sim} 0.3\,Gm/r$ 
in the O shell in hydrostatic equilibrium,
 the correlation between strong convection and
low entropy in O shells is therefore expected;
strong convection and low shell entropy are a reflection
of the same physical reason, i.e., a smaller core
size, stronger gravity, and higher temperature 
in the O shell.

\begin{figure*}
\includegraphics[width=\linewidth]{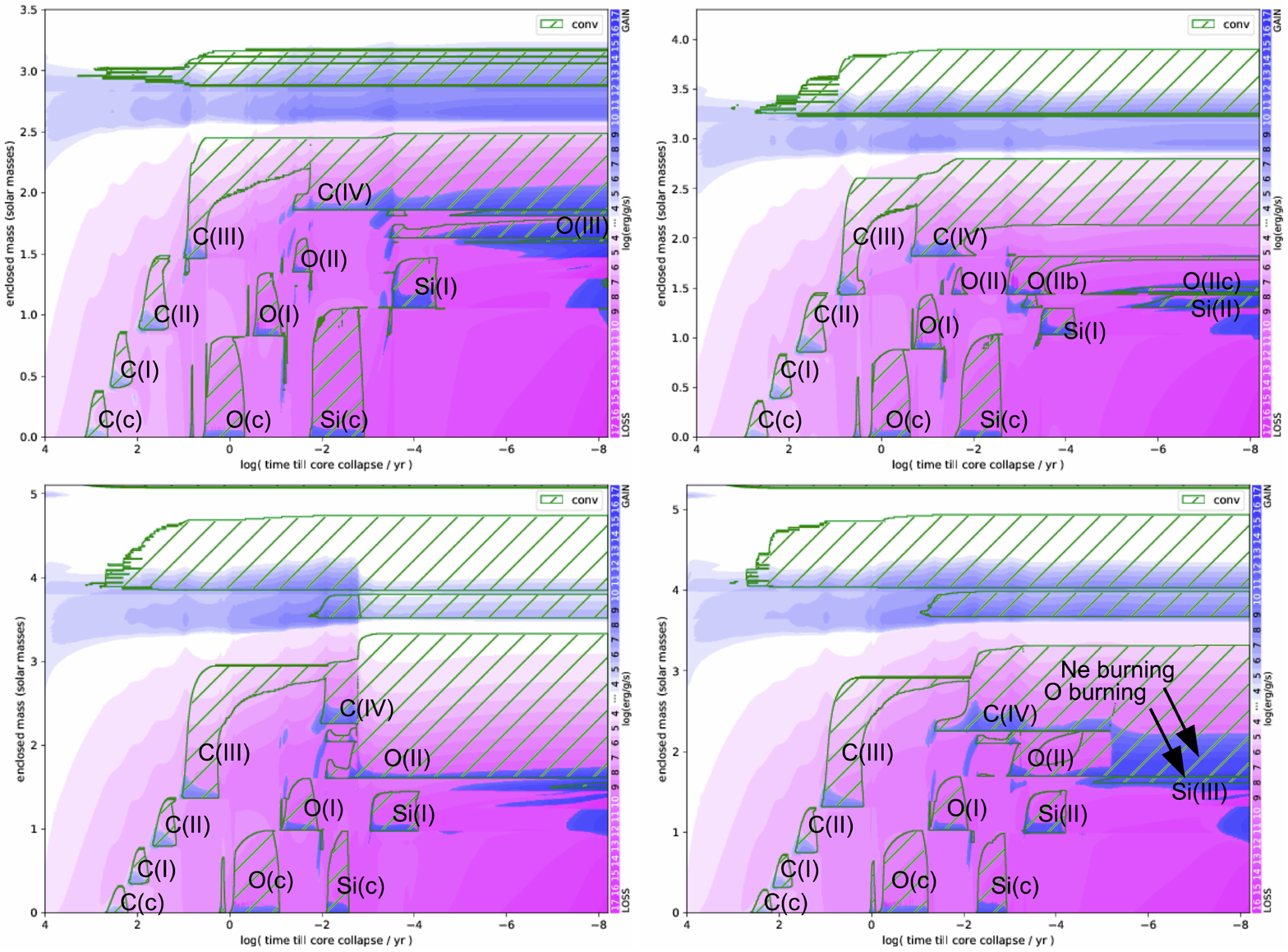}
\caption{Kippenhahn diagrams of the last $10,000\,$yr of the evolution
  of massive stars, illustrating the transition to a different shell
  configuration above $\mathord{16} \,\Msun$. Convective regions
  (hatched) driven by Si, O, Ne, or C burning are indicated explicitly
  (except for very thin shells) with Roman numerals denoting the
  various shell burning episodes and ``(c)" denoting convective core
  burning.  Colours indicate the net energy generation/cooling rate.
  At low mass, the final O shell typically remains separated from the
  C/Ne shell regardless of whether convective burning in the C/Ne
  shell remains active until collapse ($14.4 \,\Msun$ model, top left)
  or shuts off ($15.4 \,\Msun$, top right).  At higher mass, the O and
  C/Ne shell often merge ($17.1 \,\Msun$, bottom left; $17.57
  \,\Msun$, bottom right). Typically, the outer shell in the merger
    is a partially burned C shell with high Ne mass fraction as
    depicted here. If the merger occurs early ($17.1 \,\Msun$), there
  is sufficient time to mix the merged shells and largely deplete Ne
  or C so that the convection is only driven by the burning of O and
  Mg at the bottom.  For late mergers ($\mathord{\sim} 10
  \,\mathrm{min}$ for $17.57 \,\Msun$), there is insufficient time to
  thoroughly mix and deplete Ne or C, so that strong O and Ne burning
  can occur simultaneously in the same convective shell at the
  presupernova stage, though at different locations within that shell.
\label{fig:kippenhahn2}}
\end{figure*}

\begin{figure}
\includegraphics[width=\linewidth]{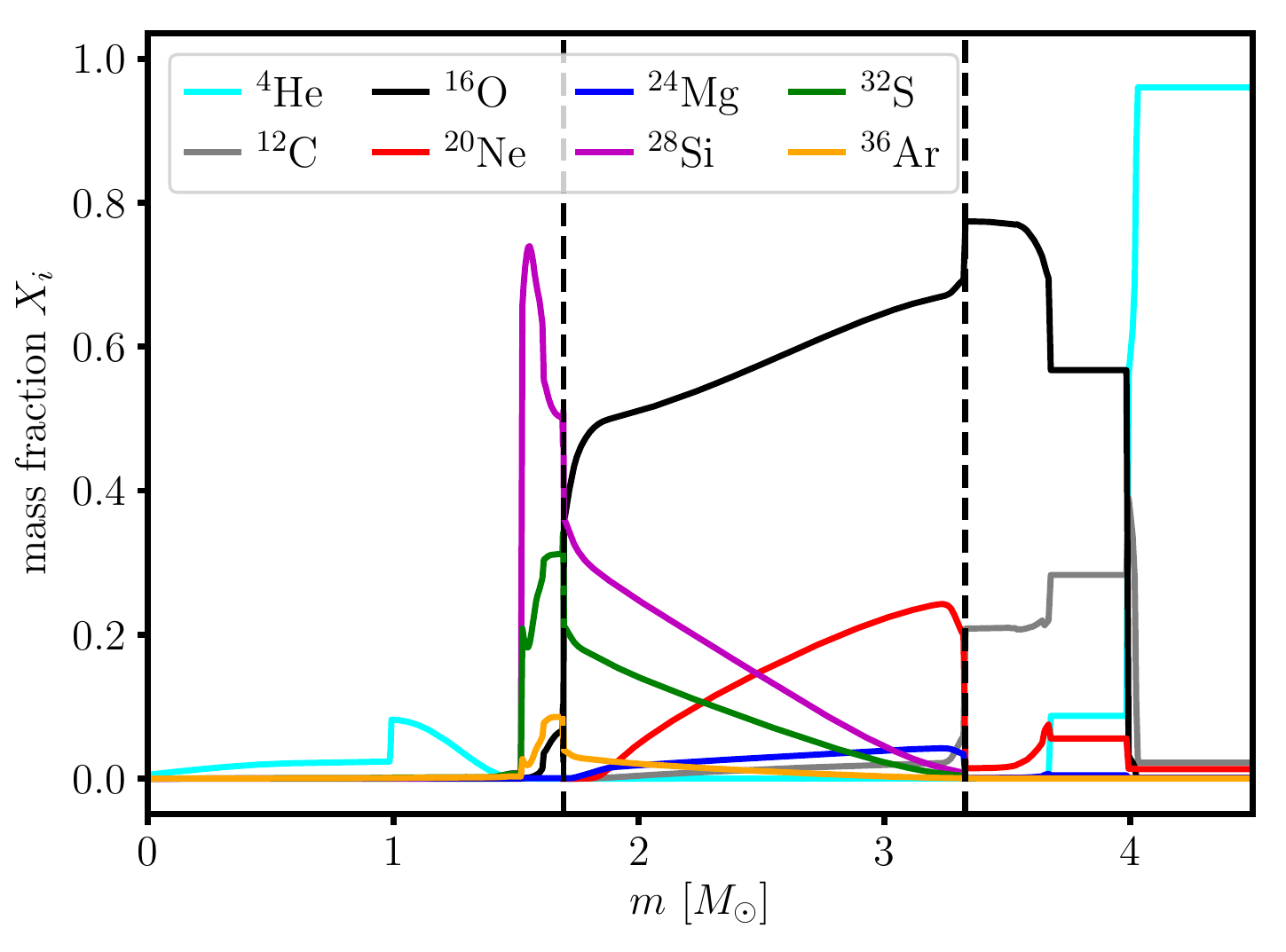}
\caption{Mass fractions of selected elements at collapse as a function
  of mass coordinate in a $17.57\,\Msun$ model with a shell merger
  about 10 minutes before collapse.  Dashed vertical lines denote
    the boundaries of the merged shell.  The outer part of the shell
    is still Ne-rich, indicating that there is insufficient time to
    homogenise the shell.  Ne is burned as it is mixed downwards, and
    any Ne is consumed before it reaches the inner shell boundary.
    Towards the inner shell boundary, O burning occurs, which is
    reflected by the steeper gradient of the O mass fraction in this
    region.
\label{fig:Xm}}
\end{figure}

\subsubsection{Role of Shell Mergers}
\label{sec:merger}

Very few other features in the landscape of convective Mach numbers
and shell geometries admit a simple explanation. One notable exception
concerns the increasing prevalence of thick convective O shells above
$\gtrsim 15 \,\Msun$. This is related to the increasing prevalence of
mergers of the O, Ne, and C shells in this mass range. The tendency
towards high O shell entropy in this region makes it relatively likely
that the buoyancy jump between the O shell and depleted or active C/Ne
shells shrinks to zero during the lifetime of the third oxygen shell,
which will lead to very massive merged convection zones (often more
than $1 \,\Msun$), as illustrated in a sequence of Kippenhahn diagrams
in Figure~\ref{fig:kippenhahn2} for progenitors of $15.4 \,\Msun$ (no
merger), $17.1 \,\Msun$, and $17.57 \,\Msun$ (merger of O and C
burning shells).  Depending on the depletion of the C/Ne shell and the
time of the merger, a rather complicated interplay of several burning
processes within one deep shell can occur.  Especially for late shell
mergers, convective mixing is no longer sufficiently effective to
homogenise the merged layers into a ``pure'' O burning shell. Instead,
one can have a situation where the Ne produced in the C shell
burns vigorously as it is mixed downwards, while O burning at the base
of the shell is also active (cp.\ the composition of the model at
  collapse shown in Figure~\ref{fig:Xm}). Sometimes the
volume-integrated nuclear energy generation rate from Ne burning
becomes comparable to that of O burning in such merged shells,
especially if the temperature at the base of the shell is relatively
low.\footnote{Even in these cases O burning still occurs and that Ne
  is fully depleted at the base of the shell, which justifies
  classifying them as O shells.}  Progenitors with simultaneous O and
Ne burning in the same shell at the pre-supernova stage are among the
ones that exhibit the highest convective Mach numbers and the largest
eddy scale. These cases are not rare; they comprise 40\% of our
progenitor models between $16\,\Msun$ and $26 \,\Msun$.

An earlier merger (i.e.\ much more than a few turnover times before
collapse) as in the $17.1 \,\Msun$ model can have a a very different
effect: Here the merged shell has sufficient time to expand due to the
energy release from the rapid burning of C and Ne from the outer shell
and then maintains a relatively low temperature at its base.  As a
result, nuclear energy generation (mainly from the burning of O and
Mg) and convection slow down, and the shell also avoids being fully
mixed. Such a structural adjustment process is responsible for another
part of the cases with low convective Mach numbers in the O shell
above $15.4\, \Msun$.

\section{Impact on the Supernova Explosion Mechanism}
\label{sec:impact}

\begin{figure}
\includegraphics[width=\linewidth]{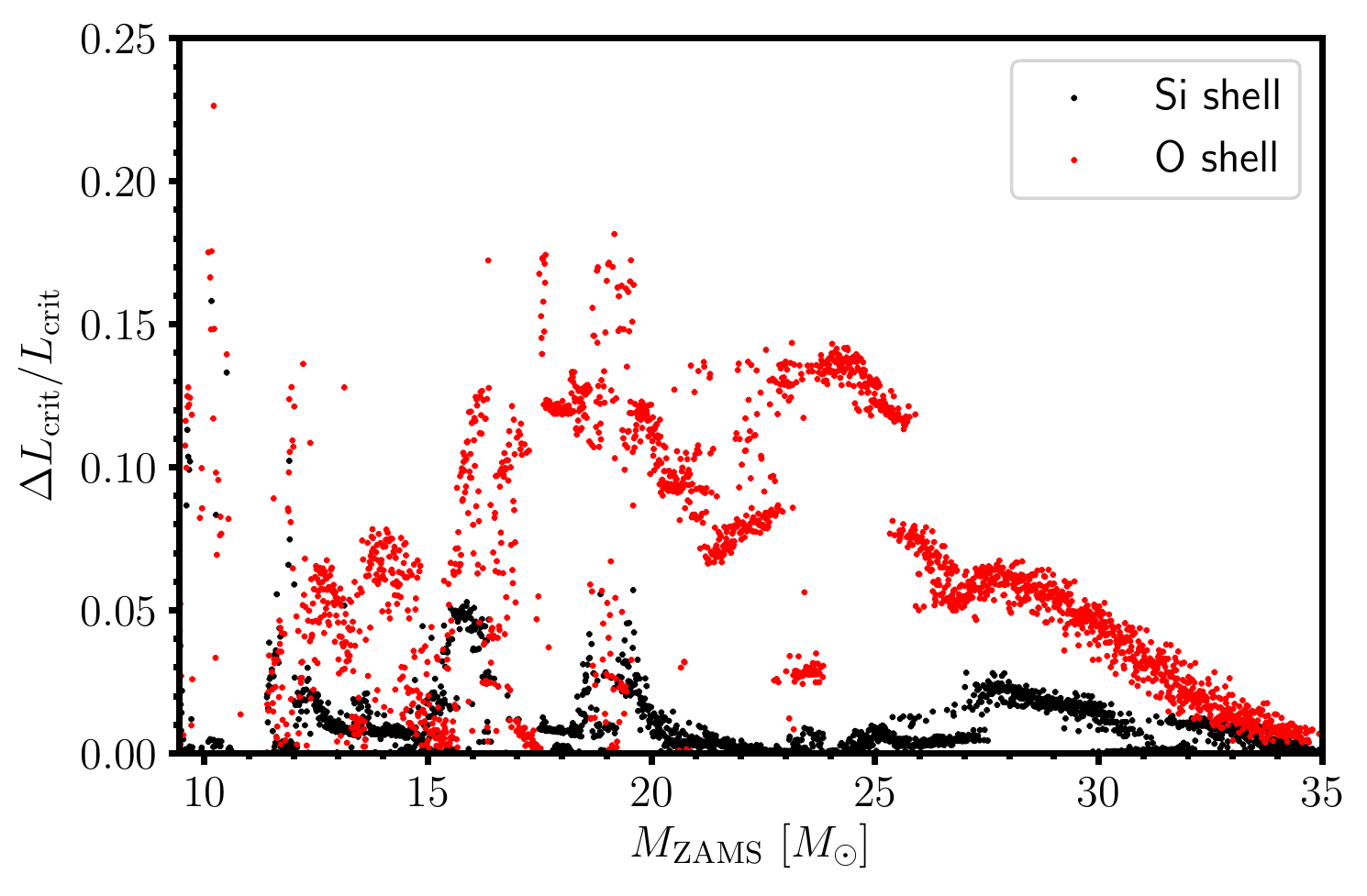}
\caption{Estimated reduction
$\Delta L_\mathrm{crit}/L_\mathrm{crit}$ of the critical luminosity for neutrino-driven explosions due to infalling perturbations in the Si shell (black) and O shell (red) according to Equation~(\ref{eq:delta_lcrit}).  The expected effect of perturbations on shock revival is strongest for perturbations in the O shell roughly between $16 \,\Msun$ and $26 \,\Msun$, and is typically small for convective perturbations in the Si shell. 
\label{fig:ma_over_l}}
\end{figure}

Although a comprehensive and accurate quantitative theory for the
interaction of perturbations in collapsing shells with the stalled
accretion shock in the supernova core is still not available,
numerical \citep{mueller_15a} and analytic
\citealp{abdikamalov_16,mueller_16c}; Huete, Abdikamalov \& Radice, in preparation) studies have established general
trends: Both high Mach numbers and small angular wave numbers in the
convective shell in question are required to boost the prospects for
neutrino-driven explosions. Based on analytic
considerations and the numerical parameter study
of \citet{mueller_15a}, \citet{mueller_16c} estimated
that infalling perturbations reduce
the critical luminosity $L_\mathrm{crit}$
\citep{burrows_93} required for a neutrino-driven
runaway by
\begin{equation}
\label{eq:delta_lcrit}
\frac{\Delta L_\mathrm{crit}}{L_\mathrm{crit}}\sim 
0.47 \frac{\mathrm{Ma}}{\ell \eta_\mathrm{acc} \eta_\mathrm{heat}}.
\end{equation}
where $\eta_\mathrm{acc}$ is the efficiency factor for the conversion
of accretion energy into electron-flavour neutrino luminosity and
$\eta_\mathrm{heat}$ is the neutrino heating efficiency (ratio of total
neutrino-heating rate and electron flavour luminosity).  For typical
values of $\eta_\mathrm{heat}\sim 0.1$ and $\eta_\mathrm{acc}\sim 2$,
one obtains $\Delta L_\mathrm{crit}/L_\mathrm{crit} \sim 2.34\,\mathrm{Ma}/\ell$.  We plot this estimated reduction of the
critical luminosity due to perturbations in the Si and O shell in
Figure~\ref{fig:ma_over_l}. It should be emphasised that this
provides only a \emph{very rough indicator} for the prospects of
perturbation-aided explosions across the stellar mass range: The
theory behind Equation~(\ref{eq:delta_lcrit}) has not yet been
validated to such a degree that it could be considered as much
more than a successful fit formula \citep{mueller_17} that
captures trends in extant numerical simulations of the
perturbation-aided mechanism
\citep{couch_13,couch_15,mueller_16c,mueller_17} within a factor
of $\mathord{\sim} 2$.

Fortunately, variations in $\Delta L_\mathrm{crit}/L_\mathrm{crit}$
are nonetheless sufficiently large to identify systematic differences between
the Si and O shell and trends with progenitor mass.

The most significant finding is that O shell burning almost invariably
provides more favourable conditions for perturbation-aided explosions
than Si shell burning as $\mathrm{Ma}/\ell$ is larger in the O shell
in $91\%$ of our progenitor models.  The expected impact of
perturbations from Si burning on the critical luminosity is small;
effect sizes of $\mathord{\sim}5\%$ are only predicted for small
clusters of models around $16\,\Msun$ and $19 \,\Msun$.  Considering
that neutrino heating conditions typically do not even get near the
threshold for runaway shock expansion in sophisticated neutrino
hydrodynamics simulations prior to the infall of the O shell
\citep{summa_16,oconnor_16}, this makes it doubtful that perturbations
from the Si shell can effectively help trigger shock revival in a larger
range of models. There is an important exception, however: In massive
progenitors with high mass accretion rates onto the supernova shock,
the SASI can develop early
\citep{mueller_12b,hanke_13,couch_14b,kuroda_16b}, and even a modest
level of pre-shock perturbations might trigger stronger SASI activity
and explosions before the infall of the O shell.  Even in this
scenario, the relatively large wave numbers of the perturbations in the
Si shell remain an obstacle; the perturbations would not provide a
strong seed for the unstable SASI modes ($\ell=1$ or $\ell=2$).

Perturbations from O shell are generally of a more favourable
amplitude and scale for the perturbation-aided mechanism, but the
expected effect size on shock revival is not uniformly high across the
stellar mass range. A large reduction of the critical luminosity due
to perturbations is mostly expected in the range between
$\mathord{\sim} 16 \,\Msun$ and $\mathord{\sim}26\,\Msun$ with a
particularly large effect size around $18\,\Msun\ldots 19\,\Msun$.
There are a few exceptions with high $\mathrm{Ma}$ and low $\ell$
below $12\,\Msun$, however, with some particularly large effect sizes
for progenitors around $10\, \Msun$.  This suggests that extant
studies of the perturbation-aided mechanism based on 3D progenitor
models (\citealp{mueller_16b,mueller_17}; $18 \,\Msun$) have already
explored the most promising region in parameter space, and that the
initial perturbations in the progenitor may generally play a smaller
role in the explosion mechanism than the first simulations suggest.
Moreover, both observations of supernova progenitors
\citep{smartt_09a,smartt_09b,smartt_15} and parameterised models of
neutrino-driven supernovae
\citep{oconnor_11,ugliano_12,ertl_15,sukhbold_16,mueller_16a} indicate
that the fraction of successful explosions drops strongly above
$15\,\Msun\ldots 18 \,\Msun$. In the face of low explodability, even
the strong large-scale perturbations in progenitors up to $26 \,\Msun$
may not be sufficient to achieve shock revival in many cases.

It is interesting to note that \citet{wongwathanarat_17} speculated
about the large-scale initial perturbations as a possible explanation
for asymmetries (Si- and Mg-rich ``jets'') in the remnant of Cas~A
that are apparently unrelated to the asymmetries of the inner Ni-rich
ejecta.  The fact that we find that violent large-scale convection
modes in the O shell start to appear above $16 \,\Msun$, i.e., for
relatively high He core mass as suggested for the progenitor of Cas~A
is in line with that speculation, although simulations are needed to bear
out this hypothesis.

\section{Uncertainties in Shell Properties}

As our identification of favourable and unfavourable mass ranges for
perturbation-aided explosions triggered during the infall of the O or
Si shell is based on 1D stellar evolution models, it is subject to
uncertainties due to potential inaccuracies of MLT and the lack of a
self-consistent treatment of convective boundary mixing. We argue,
however, that the most serious uncertainties do not undermine our
most important findings.

\subsection{Prediction of Convective Velocities}
\label{sec:mlt_uncertainties}

One of the problems of the MLT prediction for the convective velocity
(Equation~\ref{eq:vconv}) lies in the use of the pressure scale
height (or a multiple thereof) as the mixing length
$\Lambda_\mathrm{mix}$.  It has been argued \citep{arnett_09} that the shell
width is a more natural choice for $\Lambda_\mathrm{mix}$ considering
that it sets the dominant scale of the convective
eddies. \citet{mueller_16c} demonstrated, however, that the use of the
pressure scale height for $\Lambda_\mathrm{mix}$ is justified for
\emph{thick} shells, because the condition of roughly uniform entropy
generation throughout the shell for quasi-stationary convection
implies that the dissipation length in the turbulent flow cannot be
much larger than the pressure scale height.  For shells thinner than a
pressure scale height, the standard choice for the mixing length in
MLT remains questionable: Since the
local value of the mixing length should be limited by
the distance to the convective \citep{boehm_67,stothers_97},
$\Lambda_\mathrm{mix}$ should always be bounded by the shell width.
Whether MLT becomes inadequate in this regime remains to be tested by
simulations, but the qualitative effect of this uncertainty on the
systematics of $\mathrm{Ma}$ can easily be estimated. Since typical
values for the pressure scale height are $r/2\ldots r/4$ in terms of
shell radius $r$, we expect that standard MLT gives reasonable
estimates for the convective Mach number for shells up to $\ell \sim
10$. Even for shells with $\mathord{\gtrsim} 10$, $v_\mathrm{conv}$
should only be overestimated by a factor $\sim(\ell/10)^{1/3}$ by
standard based on the dependence of $v_\mathrm{conv}$ on the cube root
of the dissipation length ($v_\mathrm{conv} \sim
(\dot{\epsilon}_\mathrm{nuc} \Lambda_\mathrm{mix})^{1/3}$;
\citealp{arnett_09,mueller_16c,jones_17}).  This implies that for the majority of O
shells and the most violent Si shells, we do not expect a large error
in $\mathrm{Ma}$.  Moreover, if the convective velocities are smaller
than predicted by standard MLT, this would only accentuate our main
findings, i.e., convective perturbations in the Si shell, or in the O
shells of progenitors with $\mathord{\sim}15\,\Msun$, would be even
less likely to significantly boost neutrino-driven explosions.

The MLT predictions may require further study by means of 3D
simulations in the case of shells with simultaneous O and Ne burning,
however.  That the aftermath of shell mergers may lead to
  interesting nucleosynthesis has already been pointed out by
  \citet{ritter_17}, but the flow dynamics itself also merits
  attention in its own right since the burning and the convective
flow may depend sensitively on how Ne is mixed into the O shell.  This
is slightly reminiscent of a regime encountered in simulations of
proton ingestion \citep{stancliffe_11,herwig_11,herwig_14} where the
burning of the entrained material can be dynamically relevant and lead
to global oscillations and the establishment of a new convective
boundary \citep{herwig_14}. Although the situation is somewhat
different inasmuch as the convective zones with simultaneous O and Ne
burning arise from shells with a vanishing buoyancy jump, a better
exploration of these shell mergers is clearly called for.

Another potential problem concerns the breakdown of a quasi-stationary
balance between nuclear energy generation, buoyant driving, and
turbulent dissipation at the point when the time scale for variations
in $\dot{\epsilon}_\mathrm{nuc}$ becomes shorter than the convective
turnover time immediately before collapse \citep{mueller_16c}.
\citet{mueller_16c} demonstrated, however, that time-dependent MLT
still captures the ``freeze-out'' of convection before the onset of
collapse rather well, though it can somewhat overestimate the maximum
convective velocity within a shell. 
Again, this is unlikely to affect
our central finding of stronger convection in the O shell compared to
the Si shell: As the contraction of the shells outside the Fe core in
the last few seconds before collapse is non-homologous, changes in the
temperatures and burning rates of Si shell are generally faster than
in the O shell, and \textsc{Kepler} is thus more likely to
overestimate the convective velocities in Si shells where convection
has already undergone freeze-out. By the same token,
it is also unlikely that the ignition of unburnt Si shell
can trigger strong convection during collapse. Even though
the available energy from the fusion of
unburnt Si during collapse is often sizable
($\mathord{\sim} 10^{49}\, \mathrm{erg}$), the collapse time
would simply not be sufficient to allow sustained convection
to develop and reach the non-linear regime; this would require
convection to grow faster than on a dynamical time-scale
(which is the time-scale of the collapse). Moreover, we verified
that the general dependence of $\mathrm{Ma}$ on progenitor mass
does not change substantially if we consider the average convective
velocity instead of the maximum convective velocity, which is further evidence that
the systematic variations in $\mathrm{Ma}$ are very robust.

\subsection{Uncertainties due to Convective Boundary Mixing}
Compared to the uncertainties due to the use of time-dependent MLT in
our stellar evolution models, it is more difficult to estimate the
effect of changes in the progenitor structure due to uncertainties in
convective boundary mixing. We consider it likely that convective
boundary mixing does not change the overall trend towards more massive
and extended Fe and Si cores and higher Fe core entropy in more
massive progenitors, which is ultimately responsible for the
decreasing trend in $\mathrm{Ma}$. 

It is, however, worth pointing out that the systematics in
$\mathrm{Ma}$ could imply that shell growth by entrainment is stronger
in low-mass progenitors and could therefore help to establish wider
shells and larger convective structures in the O shells with the
highest convective Mach numbers (mostly by entraining material
from the C shell since the upper boundary is generally 
softer, see \citealp{cristini_17}). In the relevant regime, the
entrainment velocity $v_\mathrm{entr}$ (i.e., the velocity at which
the convective boundary moves) is typically assumed to be some power
law in terms of the convective velocity and the bulk Richardson number
$\mathrm{Ri}_\mathrm{b}$
\citep{fernando_91,strang_01,meakin_07,cristini_16},
\begin{equation}
v_\mathrm{entr}=
A v_\mathrm{conv} \mathrm{Ri}_\mathrm{b}^{-B},
\end{equation}
where the coefficients $A$ and $B$ are still subject to debate and may
be somewhat problem-dependent. In the case of a density discontinuity
with a relative density jump $\Delta \rho/\rho$, $\mathrm{Ri}_\mathrm{b}$
can be defined as
\begin{equation}
  \mathrm{Ri}_\mathrm{b}=
  \frac{\Delta \rho}{\rho} \frac{g l}{v_\mathrm{conv}^2}
\end{equation}
in terms of $v_\mathrm{conv}$, the local gravitational acceleration
$g$, and the characteristic horizontal length scale of convection $l$,
which we can roughly identify with $\Delta r$. 
With $c_\mathrm{s}^2\propto Gm/r$ and $l \propto r/\ell$, we find
\begin{equation}
\mathrm{Ri}_\mathrm{b} \propto \ell^{-1} \mathrm{Ma}^{-2}.
\end{equation}
In other words, the combination of high $\ell$, $v_\mathrm{conv}$, and
$\mathrm{Ma}$, which is found for O shells in progenitors below $15
\,\Msun$ is most conducive to shell growth by entrainment. 
This suggests that the structure of progenitors in this mass range
could be more seriously affected by convective boundary mixing, and
that the O shells in these stars may be wider than estimated on the
basis of our models.
Considering that $\mathrm{Ri}_\mathrm{b}$ varies considerably during
the lifetime of shells and that the proper definition of quantities
like $l$ and $\delta \rho/\rho$ is beset with ambiguities
\citep{cristini_16}, we must refrain from investigating the behaviour
of $\mathrm{Ri}_\mathrm{b}$ more systematically at this stage.

\section{Conclusions}
\label{sec:conclusions}

We investigated the properties of convective shells in supernova
progenitors at the onset of collapse with a view to the recent
scenario of ``perturbation-aided'' supernova explosions
\citep{couch_13,couch_15,mueller_15a,mueller_17}.  Our survey of
convective shells in a fine grid of 2,355 solar metallicity
current single-star progenitor models computed with the stellar
evolution code \textsc{Kepler} \citep{weaver_78,heger_10} as first
presented in \citet{mueller_16a} revealed that the conditions for
perturbation-aided explosions are very non-uniform across the mass
range of supernova progenitors and between the O and Si burning
shells.

The most favourable conditions for perturbation-aided explosions,
viz.\ high convective Mach numbers and large-scale convective modes,
are encountered in the O shells of progenitors with ZAMS masses
between $\mathord{\sim} 16 \,\Msun$ and $\mathord{\sim} 26 \,\Msun$.
An important structural reason for this is the high prevalence of
mergers between the third O shell and the C or Ne shells above $16
\,\Msun$, which leads to the formation of deep and massive O shells
prior to collapse. Such mergers often occur shortly before collapse.
In $40\%$ of our progenitors between $16 \,\Msun$ and $26 \,\Msun$,
the merger is still in progress at the onset of collapse in the sense
that Ne is still not completely depleted in the shell and continues to
burn while being mixed downward.

Although the highest convective Mach numbers of up to $\mathord{\sim}
0.3$ are found in the O shells of low-mass progenitors around $12\,\Msun$, massive shells prone to global, i.e., low-$\ell$, convective
  modes are not very prevalent in this regime.
Based on analytic arguments for the effect of perturbations
on shock revival \citep{mueller_16c}, we expect
convective seed asphericities to play a considerably less
important role
than above $16 \,\Msun$, although the reduction
of the critical luminosity for explosion due to perturbations
may still be in the range of $\mathord{\sim}5\%\ldots 10\%$.

Favourable conditions for perturbation-aided explosions are rarely
encountered in the Si shell. High convective Mach numbers $\sim 0.15$
and medium-scale convective modes with $\ell <10$ are only found
in a small fraction of progenitors, which mostly cluster around ZAMS
masses of $16 \,\Msun$ and $19 \,\Msun$.

Several conclusions regarding the viability and further investigation
of perturbation-aided explosions can be drawn from these findings.
Although supernova simulations based on 3D initial models have only
been carried out for few progenitors ($15 \,\Msun$:
\citealt{couch_15}, $18 \,\Msun$: \citealt{mueller_16b,mueller_17}), the
calculations of \cite{mueller_16b} and \citet{mueller_17} have
arguably probed the most promising regime for perturbation-aided
explosions already, and the role of perturbations in the explosion
mechanism may be more modest than appeared at first glance. Future
simulations should investigate the effect of perturbations in the
distinctively different regime revealed our current study, i.e., violent
convection in thin O shells of low-mass progenitors and, in the case
of Si shell burning, the regions around $16 \,\Msun$ and $19 \,\Msun$,
although it needs to be borne in mind that the pattern of convective
Mach numbers and eddy scales may be shifted somewhat in other stellar
evolution codes.  Furthermore, our findings provide justification for
neglecting perturbations seeded by convective silicon burning in 3D
simulations of supernova progenitors by excising the silicon core as
in \citet{mueller_16c}.

Since our study is based on spherically symmetric stellar evolution
models with a time-dependent MLT treatment of convection, our findings are beset
with some uncertainties that will need to be resolved by future
simulations of convective burning and convective boundary mixing in
the final stages of massive stars. Considering that MLT and general
principles of scale selection in convection are in good agreement with
detailed 3D simulations of convective shell burning
\citep{arnett_09,mueller_16b}, we do not expect fundamental changes.
In particular, the finding that O shell burning generally provides
more favourable conditions for perturbation-aided explosions than Si
burning is likely to remain robust. A more detailed investigation of
3D effects is mostly warranted in two regimes, namely for low-mass
progenitors with thin oxygen shells, which should be particularly
prone to strong entrainment because of the high convective Mach
numbers, and shells with simultaneous O and Ne burning in massive
progenitors, for which MLT may not adequately describe the mixing of
the merged shells on dynamical time scales and its interaction with
the burning.

\section*{Acknowledgements}
We acknowledge fruitful discussions with R.~Hirschi, H.-Th.~Janka,
S.~Sim, and S.~Woosley.  This work was supported by the Australian
Research Council through an ARC Future Fellowships FT160100035 (BM)
and Future Fellowship FT120100363 (AH) and by STFC grant ST/P000312/1
(BM).  This material is based upon work supported by the National
Science Foundation under Grant No. PHY-1430152 (JINA Center for the
Evolution of the Elements).

\bibliography{paper}

\label{lastpage}

\end{document}